\documentclass[reprint,amsmath,amssymb,aps,pra,nofootinbib]{revtex4-1}
\usepackage{graphicx}
\usepackage{bm}
\usepackage{amsmath}
\usepackage{dsfont}
\usepackage{bbold}
\usepackage{hyperref}

\def\ba{\begin{equation}\begin{array}{c}}
\def\ea{\end{array}\end{equation}}
\def\be{\ba\displaystyle}
\def\ee{\ea}

\newcommand{\tr}{{\rm tr}}

\begin{document}

\preprint{APS/123-QED}

\title{Quantum speed limits for an open system \\
 in contact with a thermal bath}

\author{N. Il'in$^1$,~~ A. Aristova$^2$,~~O. Lychkovskiy$^{1,3}$}

\affiliation{$^1$ Skolkovo Institute of Science and Technology, Skolkovo Innovation Center 3, Moscow  143026, Russia}
\affiliation{$^2$ Faculty of Physics and Earth Sciences, Leipzig University,
Linn\'{e}strasse 5, Leipzig, 04103 Germany}
\affiliation{$^3$ Department of Mathematical Methods for Quantum Technologies, Steklov Mathematical Institute of Russian Academy of Sciences, Gubkina str., 8, Moscow 119991, Russia}

\begin{abstract}
We prove fundamental rigorous bounds on the speed of quantum evolution for a quantum system coupled to a  thermal bath. The bounds are formulated in terms of expectation values of few-body observables derived from the system-bath Hamiltonian. They do not rely on the Markov approximation and, as a consequence, are applicable beyond the limit of weak system-bath coupling.
\end{abstract}

\maketitle

\section{Introduction} A Quantum Speed Limit (QSL) is an upper bound on the speed of evolution of a quantum state. The first QSL was put forward by  Mandelstam and Tamm (MT) \cite{mandelstam1945uncertainty} (see also \cite{fleming1973unitarity}), who enquired into a rigorous formulation of the time-energy uncertainty relation. A large variety of QSLs have been discovered since then, as reviewed e.g. in \cite{pfeifer1995generalized,Dodonov_2015,frey2016quantum,deffner2017quantum}. QSLs have conceptual importance as rigorous analogs of the quantum uncertainty principle in the time-energy domain. They emerge in various branches of quantum science, including optimal control theory \cite{caneva2009optimal},  quantum resource theory \cite{Campaioli_2022}, and abstract quantum information theory~\cite{deffner2020quantum}.  QSLs are deeply interrelated with the orthogonality catastrophe, adiabatic conditions, and adiabatic quantum computations \cite{lychkovskiy2017time,il'in2020quantum,lychkovskiy2018necessaryJRLR,kieu2019class,Suzuki_2020_Performance,fogarty2020orthogonality,zhu2022orthogonality}. They bound ultimate performance of classical and quantum computers \cite{margolus1998maximum,lloyd2000ultimate,svozil2005maximum,santos2015superadiabatic}, heat engines \cite{del2014more,abah2017energy}, thermometers  \cite{Campbell_2018_Precision,sekatski2021optimal} and even batteries \cite{Campaioli2017enhancing,Allan_2021_Time-optimal}.

QSLs can be particularly useful when dealing with many-body dynamics: They promise simple estimates on the rate of change of a quantum state where addressing this change exactly is not feasible. Importantly, to fulfill this promise, a QSL should be at least finite in the thermodynamic limit. Unfortunately, this is often not the case for the MT QSL and several other popular QSLs, as observed e.g. in \cite{Bukov_2019_geometric}. However, several more recent QSLs do meet this requirement \cite{Mondal_2016_quantum,il'in2020quantum,Il'in_2021_Quantum,Shiraishi_2021_Speed,albeverio_2022_Quantum,Albeverio_2022_Optimal,Hamazaki_2022_Speed}.

The goal of the present paper is to derive {\it practical} QSLs for an open quantum system coupled to a large thermal bath. We will detail what can be considered to be  ``practical'' in Section \ref{sect: setup}. At this point, we just stipulate that a practical QSL should, at least, (i) be finite in the limit of an infinitely large bath and (ii) avoid expectation values of nonlocal observables.

The above goal has been pursued previously in refs.~\cite{delCampo2013quantum,Kobayashi_2020_Quantum}. In these articles, practical QSLs have been derived in the setting where the open system dynamics is described by the Lindblad equation. The latter has its applicability limits, most importantly -- the assumption of Markovianity that is usually justified in the limit of a weak system-bath coupling \cite{Rivas_2012_Open,Mozgunov_2020_Completely,Trushechkin_2021}.

In the present paper we take a different approach -- we  derive practical QSLs based on the underlying composite Hamiltonian that describes the system, the bath and the system-bath interaction. As a consequence, the derived QSLs are universally valid for arbitrary system-bath couplings.

The rest of the paper is structured as follows.
After setting the stage in Section \ref{sect: setup}, we present  in Section \ref{sect: MDS QSL}  a QSL  by  Mondal,  Datta, and Sazim (MDS) \cite{Mondal_2016_quantum} which is the starting point for our study. This QSL typically remains finite in the limit of a large bath; however, it involves, in general, nonlocal observables and thus is not practical as such. We propose three ways to reduce it to more practical bounds. The most straightforward one is presented in Section \ref{sect: straightforward QSL}. In Section \ref{sect: T-trick} we develop a more refined approach that explicitly exploits the special structure of the thermal state of the bath.  In Section \ref{sect: noninteracting bath} we derive practical QSLs for   baths consisting of noninteracting bosons or spins. In Sections \ref{sect: example} and \ref{sect: example 2} we consider two illustrative examples.  Most proofs are relegated to Appendix~\ref{sect: proofs}. In Appendix~\ref{appendix B} we present improvements of the thermal adiabatic condition derived in ref. \cite{il'in2020adiabatic} and  a QSL derived in ref. \cite{Il'in_2021_Quantum}. These improvements are based on mathematical insights gained in the present work.

\section{Setting the stage \label{sect: setup}}

We consider the dynamics of a quantum system with the Hamiltonian $H^S_t$ that is coupled to the
external bath with the Hamiltonian $H^B$ via the coupling term $H^{\mathrm{int}}_t$. The total Hamiltonian of the system and the bath is therefore
\be\label{H}
H_t=H^S_t+H^{\mathrm{int}}_t+H^B.
\ee
Note that the system Hamiltonian and the system-bath coupling can be time-dependent, while the bath Hamiltonian does not change with time.

The  quantum state of the system and the bath  is described by the density matrix $\rho_t$ that evolves according to the von Neumann equation,
\be\label{von Neumann equation}
i\partial_t\rho_t=[H_t,\rho_t].
\ee
We assume that at $t=0$ this density matrix is initialized in a tensor product state,
\be\nonumber
\rho_0=\rho^S_{\rm in}\, \rho^B_\beta,
\ee
\be\label{Initial}
\rho^B_\beta=e^{-\beta H^B}/Z_B, \qquad Z_B=\tr \, e^{-\beta H^B}.
\ee
Here the initial state of the bath,  $\rho^B_\beta$, is thermal, while the initial state of the system,  $\rho^S_{\rm in}$, is arbitrary (mixed or pure).

Thermal baths are typically large. Therefore, we require that a practical  QSL  does not diverge in the thermodynamic limit of the bath.

The second practical requirement is to avoid expectation values of nonlocal\footnote{Here locality is understood as the absence of progressively many-body operators in the thermodynamic limit. In other words, a local operator is a few-body operator, and a nonlocal operator is a non-few-body operator. This should not be confused with the geometrical notion of locality that is not used in the present paper. For example, in a lattice spin system of $L$ sites a product of two spin operators at different sites is local (irrespective of the distance between the sites), while a product of $L/2$ spin operators is nonlocal. All Hamiltonians are assumed to be few-body. }
operators of the bath, as they can be, in general, neither calculated theoretically nor measured experimentally.\footnote{Of course, these expectation values should be calculated with respect to the initial state $\rho_0$ given by eq.~\eqref{Initial}, not the time-evolving state $\rho_t$. This is an almost trivial remark, considering our motivation to estimate the speed of evolution of a quantum state {\it without} solving a prohibitively complex many-body von Neumann equation \eqref{von Neumann equation} (see also a related discussion in ref. \cite{Kobayashi_2020_Quantum}). Note, however, that there are different agendas where a QSL can involve expectation values with respect to $\rho_t$ \cite{frey2016quantum}. In particular, this is often the case in control theory, where a desired  $\rho_t$ is {\it given} and one wishes to design a Hamiltonian that ensures this  $\rho_t$ is attained (see e.g. refs. \cite{Carlini_2006_Time-optimal,Bukov_2019_geometric}). We do not consider such cases here.}
Such expectation values can emerge from the commutation of  $\sqrt{\rho^B_\beta}$ with bath operators, as discussed below in relation to the MDS QSL~\eqref{MDS QSL}.

At the same time, we do not consider the thermodynamic limit for the system. We have in mind mostly systems of moderate sizes, such as single qubits or qudits. For this reason, we regard arbitrary operations over $\rho^S_{\rm in}$ feasible and thus practical. In particular, we will employ the square root $\sqrt{\rho^S_{\rm in}}$ and commutators between $\sqrt{\rho^S_{\rm in}}$ and operators pertaining to the system.

We remark that a number of QSLs exist that apply to an open system whose evolution is described by an explicitly given quantum channel \cite{Taddei_2013_Quantum,delCampo2013quantum,deffner2013quantum}. It should be emphasized that they are very different in scope from the bounds derived in the present paper. Obtaining a quantum channel for the system $S$ alone from the many-body Hamiltonian \eqref{H} is, in general, a prohibitively complex problem, therefore such QSLs can not be immediately applied in the setting under consideration.

To quantify the speed of evolution of a quantum state, one needs to define a metric on the space of quantum states.  A number of such metrics are known \cite{audenaert_2014_comparisons}. We choose the Hellinger distance defined as \cite{Luo_2004_Hellinger}
\be\label{heldist}
D(\rho_1,\rho_2) = 1-\tr (\sqrt{\rho_1}\sqrt{\rho_2}).
\ee
It is well suited for quantum state discrimination and relates by two-sided inequalities to other popular measures of quantum state distinguishability  \cite{holevo1972quasiequivalence,Luo_2004_Hellinger,audenaert_2014_comparisons}.

We note that often one requires a QSL to be {\it saturable}, which means the existence of  an initial state that saturates the corresponding bound. Typically, a state saturating a QSL is a superposition of a pair of eigenstates of the Hamiltonian, see e.g. \cite{frey2016quantum}. A distinctive feature of QSLs derived in the present paper is that they apply to a restricted set of very different  initial states of the form \eqref{Initial}.  It is therefore of no surprise that they are not saturated except trivial cases of no dynamics (i.e.  $D (\rho_0,\rho_t)=0$).


\section{Mondal-Datta-Sazim QSL \label{sect: MDS QSL}}

All QSLs derived in the present paper are based on the MDS QSL \cite{Mondal_2016_quantum} which reads

\begin{eqnarray}
D (\rho_0,\rho_t) & \leqslant&
1-\cos\left( \int_{0}^{t} \sqrt{2\,I_{t'}}   \, d{t'}\right),\label{MDS QSL}
\end{eqnarray}
where
\begin{eqnarray}
I_t&=&\frac12 \,\tr \left(- [ H^S_{t}+H^{\mathrm{int}}_{t}, \sqrt{\rho^S_{\rm in} }\sqrt{ \rho^B_\beta } ]^2 \right)\label{WY information}
\end{eqnarray}
is the Wigner-Yanase skew information \cite{Wigner_1963}. The inequality \eqref{MDS QSL} is valid for times $t\in [0,t_{\rm max}]$, where  $t_{\rm max}$ is the single root of the equation
\begin{equation}\label{tmax}
\int_{0}^{t_{\rm max}} \sqrt{2\,I_{t'}} \,d{t'} = \frac\pi2.
\end{equation}
For this time interval,  an upper bound on the skew information entails an upper bound on the Hellinger distance, i.e.  $I_t\leqslant x_t$ implies $D (\rho_0,\rho_t)\leqslant 1-\cos\left( \int_{0}^{t} \sqrt{2\,x_{t'}}\,d{t'}\right)$ for $t\in [0,t_{\rm max}]$. Therefore we will report our results in the form of upper bounds on  $I_t$. Note that whenever the Hamiltonian does not depend on time, $t_{\rm max}=\pi/\sqrt{8 I}$, where $I$ is the  Wigner-Yanase skew information that also does not depend on time.

When expanding the square of the commutator in the right hand side (r.h.s.) of eq. \eqref{MDS QSL}, one obtains a term
$
\tr \Big(\sqrt{\rho^S_{\rm in}}  \, H^{\mathrm{int}}_{t'} \sqrt{\rho^S_{\rm in}} e^{-\beta H^B/2} H^{\mathrm{int}}_{t'} e^{\beta H^B/2} \rho^B_\beta \Big)
$
that contains the operator
\begin{equation}\label{nonlocal operator}
e^{-\beta H^B/2}\, H^{\mathrm{int}}_{t'} \, e^{\beta H^B/2}.
\end{equation}
In general, this is a highly nonlocal operator \cite{Avdoshkin_2020} whose thermal expectation value can be neither effectively calculated nor measured in a realistic experiment. For this reason, the MDS QSL \eqref{MDS QSL} should be deemed impractical in the considered setting.

Our strategy is to derive bounds based on the MDS QSL \eqref{MDS QSL} that lack the above drawback. In the following three sections, we pursue three different approaches to this task.

Note that the bath Hamiltonian $H^B$ drops from the commutator in the MDS QSL \eqref{MDS QSL} (for a simple reason -- since $H^B$ commutes  with $\rho_0$, see Appendix~\ref{sect: proofs}). $H^B$ is extensive in the bath size, therefore its elimination is instrumental for keeping the MDS QSL (as well as bounds it will be reduced to) finite in the thermodynamic limit.

There are two special cases when the operator \eqref{nonlocal operator} is local. The first one is when the bath is noninteracting, which is considered in Section \ref{sect: noninteracting bath}. The second one is when the bath temperature is infinite, $\beta\rightarrow 0$. This is an instructive limit that will be considered when comparing QSLs derived in what follows.

For systems with finite Hilbert spaces, it is useful to additionally check the obtained QSLs  in a trivial limit
\begin{equation}\label{trivial limit}
\rho^S_{\rm in} \sim \mathbb{1},\qquad \beta\rightarrow 0.
\end{equation}
If the bath has a finite Hilbert space, the von Neumann equation \eqref{von Neumann equation} has a trivial solution $\rho_t=\rho_0\sim \mathbb{1}$ in this limit, and thus $D (\rho_0,\rho_t)=0$. Obviously, this result is reproduced by the MDS QSL \eqref{MDS QSL}.
Although for baths with infinite Hilbert spaces the limit can be more subtle mathematically, we still expect the same result on physical grounds.

Note that it is the full quantum state of the system and the bath whose evolution is bounded by the inequality \eqref{MDS QSL}. All other QSLs presented in what follows share this feature. Due to the contractivity of the Hellinger distance with respect to the partial trace, $ D(\tr_B \rho_0,\tr_B  \rho_t)\leq D(\rho_0,\rho_t)$ \cite{Luo_2004_Hellinger}, the same QSLs hold for the reduced state of the system alone.




\section{A simple relaxation of MDS QSL \label{sect: straightforward QSL}}

Since nonlocal operators in the MDS QSL \eqref{MDS QSL} appear due to the commutator, a straightforward way to avoid them is to get rid of the commutator altogether. This can be done by using a simple inequality

\begin{equation}
\tr \big(- [A, \sqrt{\rho} ]^2 \big)\leqslant 2\, \tr (A-c)^2\rho
\end{equation}
valid for an arbitrary real number $c$, quantum state $\rho$ and self-adjoint operator $A$. Applying this inequality to the MDS QSL \eqref{MDS QSL} we get

\begin{equation}\label{MDS QSL relaxed}
I_t
\leqslant  \tr \Big((H^{\mathrm{int}}_{t} +H^S_{t}-c_{t})^2 \,\rho^S_{\rm in}\, \rho^B_\beta \Big),
\end{equation}
A real function $c_t$ is, in general, arbitrary and can be chosen to optimize the bound. Alternatively, one can simply take $c_t=\tr (H^{\mathrm{int}}_t +H^S_t)\rho_0$ -- we expect that normally this will be a good choice. This is corroborated by considering a specific example in Section \ref{sect: example}, where this choice of $c_t$ is employed.

Let us discuss the properties of the QSL \eqref{MDS QSL relaxed}. First of all, this simple bound lacks any nonlocal operators and, instead, contains only expectation values of few-body observables. Second, as long as the interaction Hamiltonian does not diverge with the bath size (which we assume), the bound is manifestly finite in the limit of infinitely large bath. Thus the QSL~\eqref{MDS QSL relaxed} fulfills our practicability requirements.

An advantage of the QSL \eqref{MDS QSL relaxed} is its simple form. However, this simplicity comes at a cost: the bound can be quite loose, particularly at high bath temperatures. For example, this is evident in the trivial limit \eqref{trivial limit}: the correct result  $D (\rho_0,\rho_t)=0$ is not, in general, captured by the QSL~\eqref{MDS QSL relaxed}. Still, in other cases the QSL \eqref{MDS QSL relaxed} performs quite well, as will be exemplified in Section \ref{sect: example}.


We note that the QSL \eqref{MDS QSL relaxed} is similar to a relaxed version of the QSL (6) from ref. \cite{Shiraishi_2021_Speed} which reads ${\cal L} (\rho_0,\rho_t)\leq t \|H^{\mathrm{int}} +H^S\|$ \cite{Shiraishi_2021_Speed}, where ${\cal L} (\rho_0,\rho_t)$ is the Bures angle and  $\|...\|$ is the operator norm. The latter QSL, however, is not applicable when the norm is infinite, as is often the case for bosonic systems or reservoirs.

\section{ QSLs explicitly depending on temperature \label{sect: T-trick}}

Here we present two QSLs that exploit the special structure of the bath thermal state and explicitly depend on temperature. They read

\begin{align}
I_t \leqslant &
\, \frac{\beta^2}8   \tr \Big( - \big\{[H^{\rm int}_{t}, H^B ], \rho^S_{\rm in} \big\} \, [H^{\rm int}_{t}, H^B ] \, \rho^B_\beta  \Big)    \nonumber  \\
& +   \tr \big(- [ H^S_{t}+H^{\rm int}_{t}, \sqrt{\rho^S_{\rm in}} ]^2  \, \rho^B_\beta \big)
\label{QSL with T-trick}
\end{align}
and
\begin{equation}
I_t  \leqslant
\frac{1}8 \tr \big(- [H^S_{t}+H^{\rm int}_{t}, \, \beta H^B - \log \rho^S_{\rm in} ]^2 \, \rho^S_{\rm in}\, \rho^B_\beta \big).
 \label{QSL with T-trick log}
\end{equation}
In eq. \eqref{QSL with T-trick}, $\{\dots,\dots\}$ stands for the anticommutator.

We remind that the proofs are presented in Appendix~\ref{sect: proofs}. A technique developed in refs. \cite{il'in2020adiabatic,Il'in_2021_Quantum} is employed there. This technique allows one to replace commutators with  $ \sqrt{\rho^B_\beta}$ in the MDS QSL \eqref{MDS QSL} by commutators with $H^B$. This way the above bounds get rid of nonlocal operators, as desired.

Note that while $H^B$ has reappeared in the bounds, it enters the bounds only through the commutator $[H^{\rm int}_{t'}, H^B ]$. Typically, such commutator is an intensive quantity, and thus the bounds remain finite in the thermodynamic limit of the bath. This intuition is supported by considering specific models, see Sections~\eqref{sect: example} and~\eqref{sect: example 2}.


Observe that in the trivial limit \eqref{trivial limit} both inequalities \eqref{QSL with T-trick} and \eqref{QSL with T-trick log} reproduce the correct result  $D (\rho_0,\rho_t)=0$, in contrast to the QSL \eqref{MDS QSL relaxed}.


\begin{figure*}[t] 
		\centering
\includegraphics[width=\linewidth]{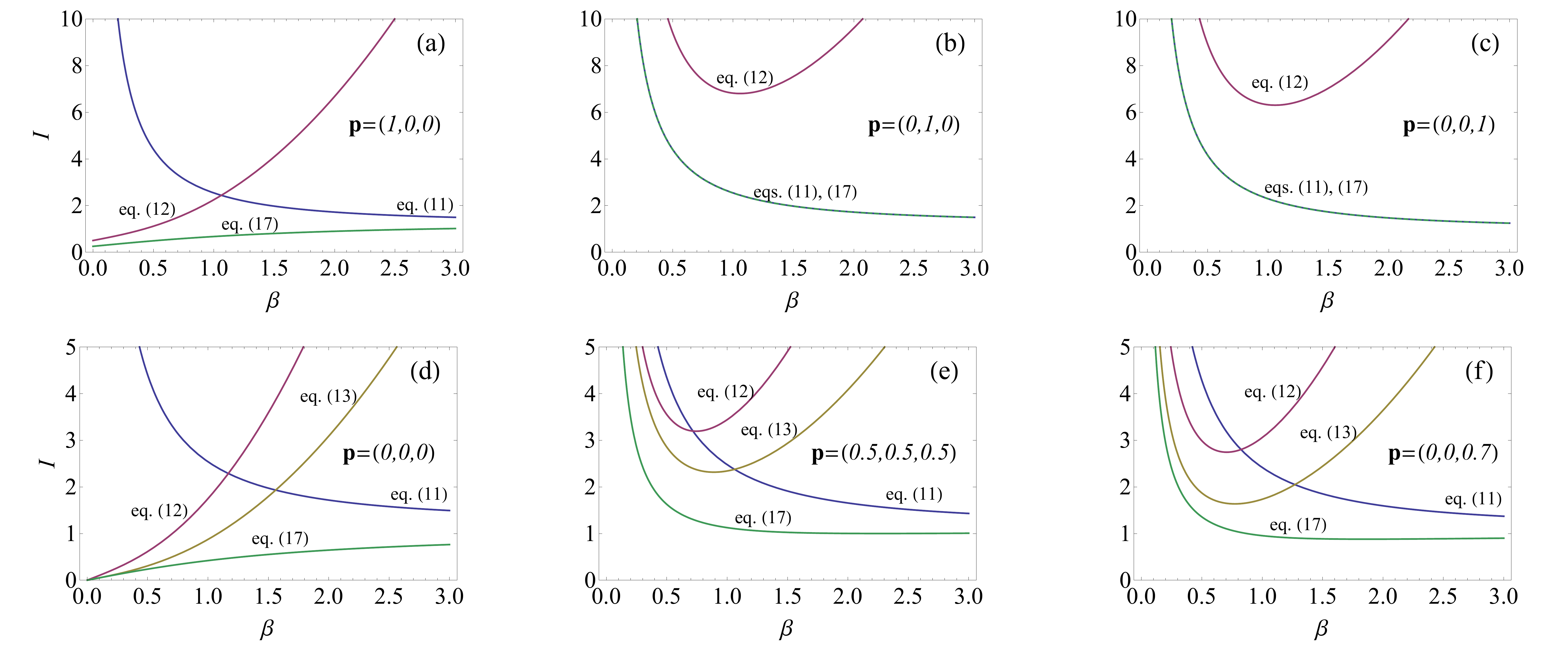}
        \caption{
        The upper bounds  \eqref{MDS QSL relaxed}, \eqref{QSL with T-trick}, \eqref{QSL with T-trick log} on the Wigner-Yanase information  \eqref{QSL noninteracting bosons} for the spin-boson model \eqref{spin-boson H} as functions of the inverse temperature $\beta$, with $\Delta=\gamma=\alpha=1$.  The initial polarization $\mathbf{p}=(p_x,p_y,p_z)$ of the spin is indicated at each plot. The upper (lower) row corresponds to the pure (mixed) initial states of the system. The bound \eqref{QSL with T-trick log} is very loose for pure initial states and thus is not shown in the upper row.
        }
		\label{fig}
\end{figure*}

\section{QSLs for noninteracting baths \label{sect: noninteracting bath}}

A noninteracting bath is a collection of noninteracting bosons, fermions, or spins. E.g. a noninteracting bosonic bath has the Hamiltonian
\be\label{HB bosons}
H^B=\sum_{k=1}^N \omega_k b_k^\dag b_k,
\ee
where $b_k$ is an annihilation operator of the $k$'th  bosonic mode, $\omega_k>0$ is the energy of the mode and $N$ is the number of modes.

For noninteracting baths, $e^{-\beta H^B/2}$ breaks into the product of one-body operators. Therefore, the operator \eqref{nonlocal operator} remains a few-body one and can be calculated explicitly. E.g. for bosons  the equalities
\begin{multline}\label{BCH}
e^{-\beta H^B/2}\,\, b_k\, e^{\beta H^B/2}=e^{\beta \omega_k /2}\,\,b_k,\\
e^{-\beta H^B/2} \, \, b_k^\dagger \, e^{\beta H^B/2}=e^{-\beta \omega_k /2}\,\,b_k^\dagger
\end{multline}
hold, and the
the computation of  \eqref{nonlocal operator} is reduced to substituting $b_k$ by $e^{\beta \omega_k /2}\,b_k$ and $b_k^\dagger$ by $e^{-\beta \omega_k /2}\,b_k^\dagger$ in  $H^{\mathrm{int}}_{t'}$.

We derive explicit forms of the MDS QSL \eqref{MDS QSL} in two cases  with noninteracting baths by calculating the Wigner-Yanase information \eqref{WY information}. In the first case, we consider a bosonic bath \eqref{HB bosons} and  a linear system-bath coupling of the form
\be
H^{\mathrm{int}}_t= \frac{\gamma}{\sqrt{N}}\sum_{k=1}^N \left(g_k S^\dagger_k b_k+ g_k^* S_k b_k^\dag\right),
\ee
where  $S_k$ are operators acting in the system's Hilbert space and $g_k$ are possibly complex coupling strengths and $\gamma$ is the dimensionless real constant. In general, $g_k$ and $S_k$  can depend on time, however, we suppress the subscript $t$ to lighten notations. The  Wigner-Yanase information \eqref{WY information} then reduces to

\begin{align}
I_t & =
 \tr_S \Bigg(
  - \frac12 [ H^S_{t}, \sqrt{\rho^S_{\rm in}} ]^2 +
 \nonumber  \\
&+\gamma^2 \sum_{k=1}^{N} \frac{|g_k|^2}{N(e^{\beta\omega_k}-1)} (e^{\beta\omega_k} S_k^\dagger  \, S_k+ S_k \,S_k^\dagger    )\rho^S_{\rm in} -   \nonumber  \\
& -\gamma^2 \sum_{k=1}^{N} \frac{|g_k|^2}{N\sinh \frac{\beta\omega_k}{2}}  S_k^\dagger \sqrt{\rho^S_{\rm in}} \, S_k \sqrt{\rho^S_{\rm in}} \Bigg).  \label{QSL noninteracting bosons}
\end{align}
One can easily check that in the trivial limit \eqref{trivial limit} this gives $I_t=0$ and hence $D (\rho_0,\rho_t)=0$ as expected.

The second case we consider here is the case of noninteracting bath of $N$ spins $1/2$,
\begin{align}\label{HB spins}
 H^B=\sum_{j=1}^N \frac{\omega_j}2 \sigma^z_j,
\end{align}
and a linear system-bath coupling of the form
\be
H^{\mathrm{int}}_t= \frac{\gamma}{\sqrt{N}}\sum_{j=1}^{N}  g_j S_j \sigma^x_j ,
\ee
where $\sigma^{x,y,z}_j$ are Pauli matrices, $S_j$ are self-adjoint operators acting in the system's Hilbert space and $\gamma$, $g_j$ are real coupling constants. In this case the   The  Wigner-Yanase information \eqref{WY information} reads
\begin{align}
I_t & =
\tr_S \Bigg(  -  \frac{1}{2} [ H^S_{t}, \sqrt{\rho^S_{\rm in}} ]^2 +
 \nonumber  \\
&+\gamma^2 \sum_{j=1}^{N} \frac{g_j^2}{N}\,  S_j^2 \rho^S_{\rm in} - \nonumber \\
&-\gamma^2 \sum_{j=1}^{N} \frac{g_j^2}{N\cosh \frac{\beta\omega_j}{2}}\,S_j \sqrt{\rho^S_{\rm in}} \, S_j \sqrt{\rho^S_{\rm in}}  \Bigg).
\label{QSL noninteracting spins}
\end{align}
Again, we verify that  $I_t=0$ in the limit \eqref{trivial limit}, as expected.

Note that while the non-negativity of the r.h.s. of the bounds \eqref{QSL noninteracting bosons},\eqref{QSL noninteracting spins} is not apparent, it is guaranteed by the non-negativity of the Wigner-Yanase information \eqref{WY information}.

\section{Example 1: spin-boson model \label{sect: example}}

A system with a noninteracting bath can serve as an instructive example for assessing and comparing the performance of QSLs derived in the previous sections. We perform such a comparison for a spin-boson model with
\begin{align}\label{spin-boson H}
H^S=\frac{\Delta}2 \sigma^z, \qquad H^B=\sum_{k=1}^N \omega_k b_k^\dag b_k, \nonumber \\
H^{\mathrm{int}}= \frac{\gamma}{\sqrt{N}}\sigma^x\sum_{k=1}^N g_k \left(b_k+b_k^\dag\right), \quad
\end{align}
where $b_k$ are boson operators, $\sigma^\alpha, \alpha=x,y,z$ are Pauli matrices, $g_k$ are constants with the dimension of energy and $\gamma$ is the overall dimensionless interaction strength. We consider the thermodynamic limit $N\rightarrow \infty$ for the bath and introduce the spectral density $J(\omega)$ that, by definition, satisfies
\be\label{thermodynamic limit}
N^{-1}\sum_k |g_k|^2 f(\omega_k) \xrightarrow{N\rightarrow \infty} \int_0^\infty d\omega J(\omega) f(\omega)
\ee
for an arbitrary smooth function $f(\omega)$. For concrete calculations, we choose a regularized Ohmic spectral density of the form
\be\label{spectral density}
J(\omega)= \omega e^{-\alpha \omega},
\ee
where $\alpha$ is a constant responsible for the high-energy cutoff.

We consider a general initial state of the spin given by
\be
\rho^S_{\rm in}=\frac{1}{2}(1+\mathbf{p}\boldsymbol{\sigma}), \quad \mathbf{p}=p\mathbf{n}, \quad |\mathbf{n}|=1, \quad p\leqslant 1.
\ee
Note that for an arbitrary function $f$
\begin{multline}
f\left(\frac{1}{2}(1+\mathbf{p}\boldsymbol{\sigma})\right)=A+B\mathbf{n}\boldsymbol{\sigma},~~~~~~~~{\rm where}\\
A=\frac{1}{2}\left(f\left(\frac{1+p}{2}\right)+f\left(\frac{1-p}{2}\right)\right),\\
B=\frac{1}{2}\left(f\left(\frac{1+p}{2}\right)-f\left(\frac{1-p}{2}\right)\right).
\end{multline}

\begin{figure*}[t] 
		\centering
\includegraphics[width=\linewidth]{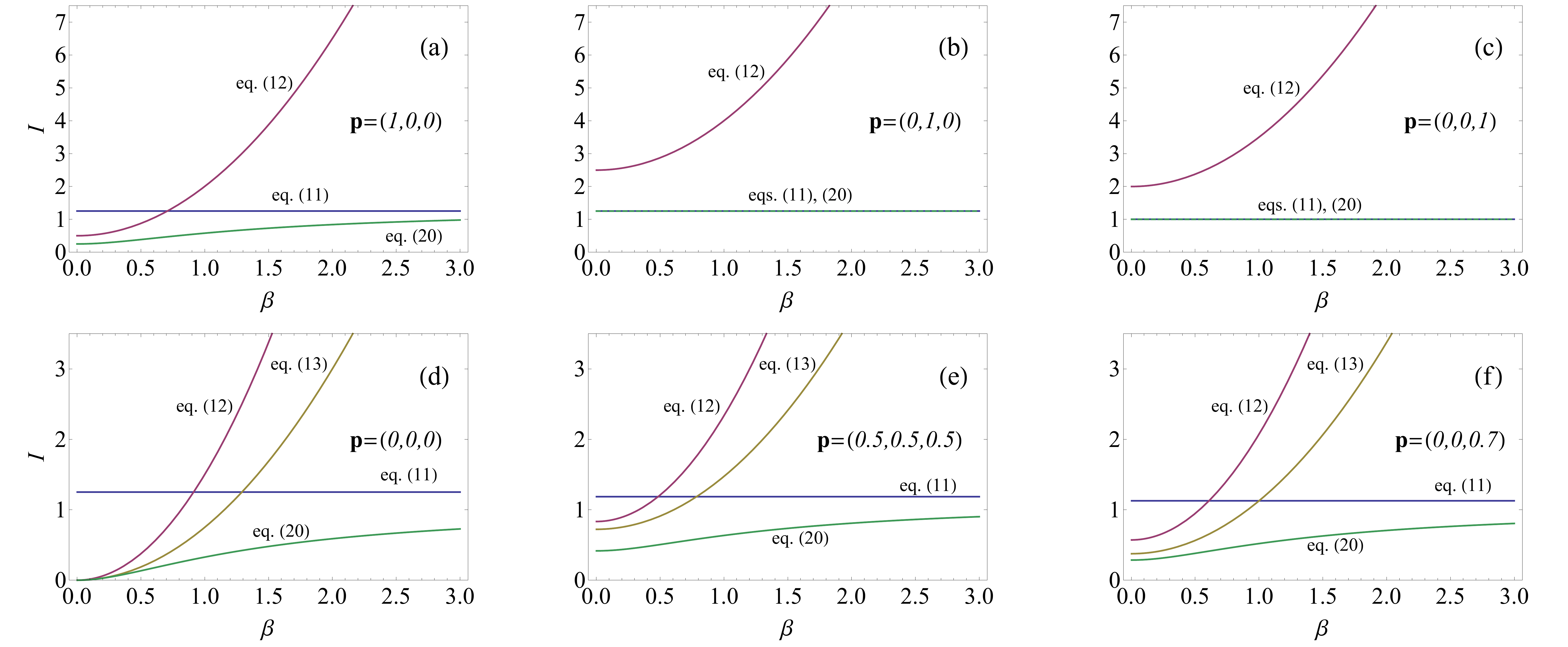}
        \caption{
        The upper bounds  \eqref{MDS QSL relaxed}, \eqref{QSL with T-trick}, \eqref{QSL with T-trick log} on the Wigner-Yanase information  \eqref{QSL noninteracting spins}  for the central spin model \eqref{central spin H} as functions of the inverse temperature $\beta$, with $\Delta=\gamma=\alpha=1$.  The initial polarization $\mathbf{p}=(p_x,p_y,p_z)$ of the spin is indicated at each plot. The upper (lower) row corresponds to the pure (mixed) initial states of the system. The bound \eqref{QSL with T-trick log} is very loose for pure initial states and thus is not shown in the upper row.
        }
		\label{fig 2}
\end{figure*}

We introduce the following notations for integrals that will appear in the QSLs:
\be\label{K}
K_\beta=\beta^2\int\limits_0^\infty J(\omega) \frac{e^{\beta  \omega}+1}{e^{\beta  \omega}-1}d\omega,
\ee

\be\label{I}
M_\beta=\beta^4\int\limits_0^\infty \omega^2 J(\omega) \frac{e^{\beta  \omega}+1}{e^{\beta  \omega}-1}d\omega ,
\ee

\be\label{L}
L_\beta=\beta^2 \int\limits_0^\infty J(\omega) \frac{e^{\frac{\beta \omega}{2}}}{e^{\beta  \omega}-1}d\omega.
\ee

The QSL \eqref{MDS QSL relaxed} reduces to
\be\label{MDS QSL relaxed example}
I_t  \leqslant
\frac{1-p_z^2}{4}\Delta ^2
+\frac{\gamma^2}{\beta^2}K_\beta.
\ee
The QSL \eqref{QSL with T-trick} reduces to
\begin{multline}\label{QSL with T-trick example}
I_t \leqslant \frac{\gamma^2}{4\beta^2}  M_\beta + 2 \Delta^2 B_p^2 (1 - n^2_z) + \\
+ 8 B_p^2 (1-n^2_x) \frac{\gamma^2}{\beta^2} K_\beta
\end{multline}
with
\be\label{Bp}
B_p^2=\frac{1}{4}\left(1-\sqrt{1-p^2}\right).
\ee
The QSL \eqref{QSL with T-trick log} reduces to
\begin{multline}\label{QSL with T-trick log example}
I_t \leqslant \frac{\gamma^2}{8\beta^2}  M_\beta + \frac{1}{8} \Delta^2 \tilde{B}_p^2 (1 - n^2_z) +\\
+ \frac{1}{2}\tilde{B}_p^2 (1-n^2_x) \frac{\gamma^2}{\beta^2} K_\beta
\end{multline}
with
\be\label{tilde Bp}
\tilde{B}_p=\ln\sqrt{\frac{1+p}{1-p}}.
\ee
Finally, the exact value of the Wigner-Yanase information  \eqref{QSL noninteracting bosons} reads
\begin{multline}\label{QSL noninteracting bosons example}
I_t = \Delta^2 B_p^2 (1 - n^2_z) +  \frac{\gamma^2}{\beta^2} K_\beta - \\
- \frac{2\gamma^2}{\beta^2} L_\beta (1 - 4B_p^2 (1-n^2_x) ).
\end{multline}
For the specific spectral density \eqref{spectral density} one can calculate $K_\beta$, $M_\beta$ and $L_\beta$ explicitly:
\begin{multline}
K_\beta=2\psi^{(1)}\left(\frac{\alpha}{\beta}\right)-\frac{\beta^2}{\alpha^2},\quad
M_\beta=2\psi^{(3)}\left(\frac{\alpha}{\beta}\right)-6\frac{\beta^4}{\alpha^4},\\
L_\beta= \psi^{(1)}\left(\frac{\alpha}{\beta} + \frac{1}{2}\right),
\end{multline}
where
$\psi^{(m)}(z)$ is the polygamma function of order $m$,
\be
\psi^{(m)}(z)=\int\limits_0^\infty\frac{t^m e^{-zt}}{1-e^{-t}}dt=\frac{d^{m+1}}{dz^{m+1}}\ln \Gamma(z),
\ee
and $\Gamma(z)$ is the Euler gamma function.

Using the above formulae,  we compare the four QSLs. Several illustrative plots are presented in Fig.~\ref{fig}. Of course, the QSL \eqref{MDS QSL} with the Wigner-Yanase information given by eq. \eqref{QSL noninteracting bosons} is never outperformed by QSLs \eqref{MDS QSL relaxed}, \eqref{QSL with T-trick}, \eqref{QSL with T-trick log}. This is because the later are different relaxations of the former. We emphasize, however, that for interacting baths the analog of eq.~\eqref{QSL noninteracting bosons} is not available, and one is left with the three remaining ones. Fig. \ref{fig} illustrates that each of them can outperform the other two for certain bath temperatures and system initial states.

One can see from Fig. \ref{fig} that for pure initial states of the system, the QSL \eqref{MDS QSL relaxed} typically outperforms QSLs \eqref{QSL with T-trick} and \eqref{QSL with T-trick log}. Furthermore, it turns out to coincide with the QSL \eqref{QSL noninteracting bosons} for initial polarizations along  $y$- or $z$-axes. This can be easily understood from eqs. \eqref{MDS QSL relaxed example}, \eqref{QSL noninteracting bosons example}. However, this superiority of the QSL \eqref{MDS QSL relaxed} is abruptly lost as soon as the initial state ceases to be pure and the bath temperature is sufficiently high, which is particularly evident by comparing plots (c) and (f) in Fig. \ref{fig}. Under the later conditions it is the QSL \eqref{QSL with T-trick log} which typically works better.

\section{Example 2: central spin model \label{sect: example 2}}

\begin{figure*}[t] 
		\centering
\includegraphics[width=\linewidth]{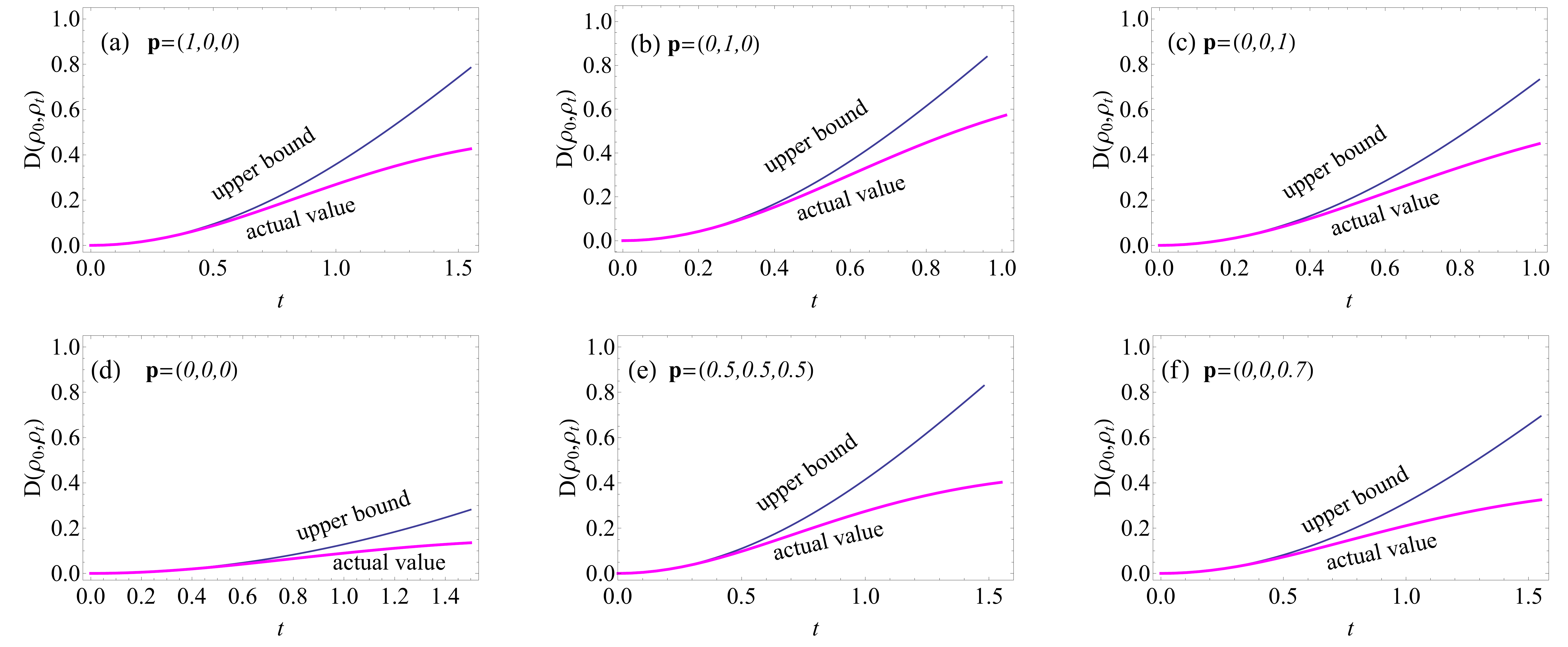}
        \caption{
        QSL  \eqref{QSL noninteracting spins} compared to the actual evolution of $D(\rho_0,\rho_t)$ for the central spin model \eqref{central spin H} with $N=10$ spins in the bath and $\Delta=\gamma=\alpha=\beta=1$.  The couplings $g_j$ and energies $\omega_j$ of the bath spins are randomly sampled from the intervals $[0.5,1.5]$ and $[0,2]$, respectively. Note different dime scales for different initial polarizations.
        }
		\label{fig 3}
\end{figure*}

As a second example, we consider a central spin model given by

\begin{align}\label{central spin H}
H^S=\frac{\Delta}2 \sigma^z, \qquad H^B=\sum_{j=1}^N \frac{\omega_j}2 \sigma^z_j, \nonumber \\
H^{\mathrm{int}}= \frac{\gamma}{\sqrt N}\sigma^x\sum_{j=1}^N g_j \sigma^x_j. \quad
\end{align}
Here a ``central'' spin $1/2$ is coupled to a bath of $N$ noninteracting spins $1/2$.

It turns out that for the central spin model the bounds  \eqref{MDS QSL relaxed}, \eqref{QSL with T-trick}, \eqref{QSL with T-trick log} and  the equality \eqref{QSL noninteracting spins} can be written in the form \eqref{MDS QSL relaxed example}, \eqref{QSL with T-trick example}, \eqref{QSL with T-trick log example}, \eqref{QSL noninteracting bosons example}, respectively, but with different expressions for $K_{\beta}$, $I_{\beta}$ and $L_\beta$:

\begin{align}
K_{\beta}&=\frac{1}{N}\beta^2\sum_{j=1}^N g_j^2 ~~~\longrightarrow \frac{\beta^2}{\alpha^2},\nonumber \\
I_{\beta}&=\frac{1}{N}\beta^4\sum_{j=1}^N g_j^2\omega_j^2 \longrightarrow 6\frac{\beta^4}{\alpha^4},\nonumber \\
L_{\beta}& =\frac{1}{N} \beta^2 \sum_{j=1}^N \frac{g_j^2}{2\cosh\frac{\beta\omega_j}{2}} \nonumber\\[10pt]
         &   \longrightarrow  \frac14 \left(
   \psi^1\left( \frac14+\frac\alpha{2 \beta}\right)
   -\psi^1\left( \frac34 + \frac\alpha{2 \beta}\right)
   \right).\label{finite N}
\end{align}
Here we provide expressions for finite $N$ along with the expressions in the thermodynamic limit. The latter ones are computed from the former ones according to eq. \eqref{thermodynamic limit} with the Ohmic spectral density \eqref{spectral density}.

In Fig.~\ref{fig 2} we compare the four QSLs in the thermodynamic limit. Among the relaxed QSLs \eqref{MDS QSL relaxed}, \eqref{QSL with T-trick}, \eqref{QSL with T-trick log}, each one can outperform the other two for certain bath temperatures and system initial states. Qualitatively, the comparative behavior of the QSLs for the central spin model is analogous to that for the spin boson model, as can bee seen from Figs. \ref{fig} and \ref{fig 2}.

Finally, we explore how tight our QSLs are. To this end we compute $D(\rho_0,\rho_t)$ numerically  for a finite bath consisting of $N=10$ spins. The  couplings $g_j$ and energies $\omega_j$ of the bath spins are picked at random from the intervals $[0.5,1.5]$ and $[0,2]$, respectively. The actual value of $D(\rho_0,\rho_t)$  is compared to the tightest QSL available for the central spin model -- the MDS QSL with the explicitly calculated Wigner-Yanase information \eqref{QSL noninteracting spins}. To compute it we use the finite-$N$ versions of expressions \eqref{finite N}. The result is shown in Fig. \ref{fig 3}. One can see that the QSL  is extremely tight at small  times but loosens as the evolution proceeds. This is a reasonable performance, given the general nature of the bounds and the inherent complexity of the underlying many-body problem.

\section{Summary and concluding remarks}

In summary, we have proved quantum speed limits  \eqref{MDS QSL relaxed}, \eqref{QSL with T-trick},  \eqref{QSL with T-trick log}  for open systems in contact with thermal baths. They constitute different relaxations of the MDS QSL \eqref{MDS QSL} \cite{Mondal_2016_quantum}.  Furthermore, we explicitly computed  the MDS bound \eqref{MDS QSL} in cases  of noninteracting bosonic and spin baths, see eqs. \eqref{QSL noninteracting bosons} \eqref{QSL noninteracting spins}. In contrast to prior knowledge, these QSLs
do not rest on Markovian or other approximations and are applicable irrespective of the strength and time dependence of the system-bath couplings. Importantly, all reported bounds are practical when applied to many-body baths, i.e. they do not diverge in the thermodynamic limit and contain only local quantities that can be calculated or measured.

The MDS QSL \eqref{MDS QSL}, \eqref{QSL noninteracting bosons}, \eqref{QSL noninteracting spins} provides the tightest bounds for noninteracting baths but is not applicable to interacting baths.   The QSLs \eqref{MDS QSL relaxed}, \eqref{QSL with T-trick},  \eqref{QSL with T-trick log} are applicable for interacting baths and  complement each other in different regimes: the QSL \eqref{MDS QSL relaxed} is typically the best for pure initial states of the system or low bath temperatures, while QSLs \eqref{QSL with T-trick} and  \eqref{QSL with T-trick log} perform better for mixed initial  states of the system and sufficiently large bath temperatures.

The bounds reported in the present paper allow one to quickly  estimate the evolution speed of open quantum systems coupled to thermal reservoirs, particularly in the non-Markovian regime where solving the dynamical equations exactly can pose considerable difficulties. This can be useful in various applications, including the dissipative error correction \cite{Cohen_2014,reiter2017dissipative,de2022error},  dissipative entanglement generation \cite{Krauter_2011,martin2011dissipation,Cole_2022}, dissipation-enhanced quantum sensing \cite{Checinska_2015,reiter2017dissipative,Chen_2017,Raghunandan_2018},  structured reservoir engineering \cite{lutkenhaus1998mimicking,schirmer2010stabilizing,Roy-Choudhury_2015,Gonzalez_2017,basilewitsch2019reservoir} {\it etc}. Another potential application of the reported results is to provides rigorous benchmarks for various computational methods addressing non-Markovian quantum dynamics.

\section{Acknowledgements}
This work was supported by the Russian Science Foundation under the grant No 17-71-20158.


\appendix

\bigskip
\section{Proofs \label{sect: proofs}}

%
%
%
%

\subsection{Lemma \label{subsect: Lemma}}

The proof of eqs. \eqref{QSL with T-trick} and \eqref{QSL with T-trick log} employs the following auxiliary

\medskip

\noindent {\it Lemma.} For arbitrary real $x$ and $y$
\be\label{exp3}
\left(e^{-x}-e^{-y}\right)^2\leqslant\frac{e^{-2x}+e^{-2y}}{2}(x-y)^2.
\ee

\medskip

\noindent To prove the Lemma, we use the Hermite-Hadamard  inequality \cite{Hadamard_1893}
\be\label{HH}
\int_x^yf(t)dt\leqslant\frac{f(x)+f(y)}{2}(y-x)
\ee
valid for any convex function $f(t)$.
For  $f(t)=e^{-t}$, it takes the form
\be
e^{-x}-e^{-y}\leqslant\frac{e^{-x}+e^{-y}}{2}(y-x), \quad x\leqslant y.
\ee
This inequality leads to eq. \eqref{exp3} after simple algebraic manipulations.  The range of eq. \eqref{exp3} is immediately expanded to  $y\leqslant x$ on symmetry grounds.

\subsection{ Proof of eq. \eqref{QSL with T-trick} \label{subsect: proof 10}}

Note that
\begin{multline}\label{Comm_rho}
[H^S_{t'}+H^{\mathrm{int}}_{t'}, \sqrt{\rho^S_{\rm in} \rho^B_\beta} ]
=\sqrt{\rho^S_{\rm in}} \, [H^{\mathrm{int}}_{t'}, \sqrt{\rho^B_\beta }] +  \\
+ [H^S_{t'}+H^{\mathrm{int}}_{t'}, \sqrt{\rho^S_{\rm in} }] \sqrt{\rho^B_\beta}.
\end{multline}
Multiplying \eqref{Comm_rho} by its conjugate, taking the trace and applying the inequality
\be
\tr(A^\dag B+AB^\dag)\leqslant \tr A^\dag A+\tr B^\dag B,
\ee
we get
\begin{multline}\label{trace_estim}
-\tr [H^S_{t'}+H^{\mathrm{int}}_{t'}, \sqrt{\rho^S_{\rm in} \rho^B_\beta} ]^2 \leq
 -2\Big( \tr  ([H^{\mathrm{int}}_{t'}, \sqrt{\rho^B_\beta } ]^2 \rho^S_{\rm in})
+ \\
+ \tr ([H^S_{t'}+H^{\mathrm{int}}_{t'}, \sqrt{\rho^S_{\rm in} } ]^2 \rho^B_\beta) \Big).
\end{multline}
Then we plug this inequality in the MDS QSL \eqref{MDS QSL}. The second term in this inequality directly leads to the second term in eq. \eqref{QSL with T-trick}. To estimate the first term, we employ a technique similar to that in \cite{il'in2020adiabatic,Il'in_2021_Quantum}. We expand the trace in the product eigenbasis $|\mu,m \rangle=|\mu\rangle|m\rangle$, where $|\mu\rangle$ are eigenvectors of  $\rho^S_{\rm in}$ and $|m\rangle$ are eigenvectors of  $H^B$:
\begin{multline}\label{proof eq 2}
 -\tr  ([H^{\mathrm{int}}_{t'}, \sqrt{\rho^B_\beta } ]^2 \rho^S_{\rm in})
  \\
 = -\frac{1}{2}\tr  ([H^{\mathrm{int}}_{t'}, \sqrt{\rho^B_\beta } ] \, \{[H^{\mathrm{int}}_{t'}, \sqrt{\rho^B_\beta } ],\, \rho^S_{\rm in}\})\\
 = \frac{1}{2Z_B}\!\!\sum_{\mu,\nu,m,n} \!\!\!
\left(e^{-\frac{\beta}{2}E_m}-e^{-\frac{\beta}{2}E_n}\right)^2\!\!
(w_\mu+w_\nu)\big|\langle \mu,m |  H^{\mathrm{int}}_{t'} |\nu,n\rangle\big|^2,
\end{multline}
where $w_\mu$ and $w_\nu$ are eigenvalues of $\rho^S_{\rm in}$.
For the function $e^{-\beta E/2}$ the inequality \eqref{exp3} reads
\be\label{exp4}
\left(e^{-\frac{\beta E_m}{2}}-e^{-\frac{\beta E_n}{2}}\right)^2
\leqslant \frac{\beta^2}{8}(e^{-\beta E_m}+e^{-\beta E_n})(E_m-E_n)^2.
\ee
Using \eqref{exp4} we estimate \eqref{proof eq 2} as
\begin{multline}\label{proof eq 21}
 -\tr  ([H^{\mathrm{int}}_{t'}, \sqrt{\rho^B_\beta } ]^2 \rho^S_{\rm in})
  \leqslant \\
  \leqslant -\frac{\beta^2}{16} \Big(
  \tr \big([ H^{\mathrm{int}}_{t'}, H^B ]\{[ H^{\mathrm{int}}_{t'}, H^B ] ,\rho^S_{\rm in}\}  \rho^B_\beta  \big)+\\
  +\tr \big(\{[ H^{\mathrm{int}}_{t'}, H^B ],\rho^S_{\rm in}\} \rho^B_\beta\, [ H^{\mathrm{int}}_{t'}, H^B ] \,   \big)
  \Big) =\\
  =-\frac{\beta^2}8\tr \big([H^{\mathrm{int}}_{t'}, H^B ]\{[ H^{\mathrm{int}}_{t'}, H^B ] ,\rho^S_{\rm in}\}  \rho^B_\beta  \big).
\end{multline}

Combining eq. \eqref{proof eq 21} with eqs. \eqref{trace_estim} and \eqref{MDS QSL} completes the proof of the bound \eqref{QSL with T-trick}.

\subsection{ Proof of eq. \eqref{QSL with T-trick log}}

We rewrite the integrand in eq. \eqref{MDS QSL} as
\begin{multline}\label{Tr_equality}
-\tr\big([ H^S_{t'}+H^{\mathrm{int}}_{t'},\sqrt{\rho^S_{\rm in}}\sqrt{\rho^B_\beta }]^2\big)
=\\
= \sum_{\mu,\nu,m,n}|\langle \mu,m | [ H^S_{t'}+H^{\mathrm{int}}_{t'},\sqrt{\rho^S_{\rm in}}\sqrt{\rho^B_\beta }] |\nu,n\rangle|^2=\\
=Z_B^{-1}\sum_{\mu,\nu,m,n}(e^{-\frac{\beta E_n}{2}}\sqrt{w_\nu}-e^{-\frac{\beta E_m}{2}}\sqrt{w_\mu})^2 \times \\
\times |\langle \mu,m | H^S_{t'}+H^{\mathrm{int}}_{t'} |\nu,n\rangle|^2.
\end{multline}
Let us introduce a new variable
\be
\delta_\nu=-\ln w_\nu.
\ee
Applying the inequality \eqref{exp3} to
\be
e^{-\frac{\beta E_n}{2}}\sqrt{w_\nu}-e^{-\frac{\beta E_m}{2}}\sqrt{w_\mu}=e^{-\frac{\beta E_n+\delta_\nu}{2}}-e^{-\frac{\beta E_m+\delta_\mu}{2}},
\ee
we get
\begin{multline}
\left(e^{-\frac{\beta E_n}{2}}\sqrt{w_\nu}-e^{-\frac{\beta E_m}{2}}\sqrt{w_\mu}\right)^2
\leqslant \\
\leqslant \frac{1}{8}\big(e^{-\beta E_n}w_\nu
+e^{-\beta E_m}w_\mu\big)
(\beta E_n+\delta_\nu-\beta E_m-\delta_\mu)^2.\nonumber
\end{multline}
Plugging this into eq. \eqref{Tr_equality}, we get
\begin{multline}\label{Tr_equality2}
-\tr\big([ H^S_{t'}+H^{\mathrm{int}}_{t'},\sqrt{\rho^S_{\rm in}}\sqrt{\rho^B_\beta }]^2\big)
\leqslant \\
\leqslant \frac{1}{8}Z_B^{-1}\sum_{\mu,\nu,m,n}\left(e^{-\beta E_n}w_\nu+e^{-\beta E_m}w_\mu\right)\times\\
\times(\beta E_n+\delta_\nu-\beta E_m-\delta_\mu)^2|\langle \mu,m | H^S_{t'}+H^{\mathrm{int}}_{t'} |\nu,n\rangle|^2 \\
=-\frac{1}{4}\tr([H^S_{t'}+H^{\mathrm{int}}_{t'},\beta H^R-\ln \rho^S_{\rm in}]^2 \, \rho^S_{\rm in}\rho^B_\beta).
\end{multline}
Plugging this inequality to the MDS QSL \eqref{MDS QSL}, we obtain the bound \eqref{QSL with T-trick log}.

\bigskip
\section{Improvements of the thermal QSL \cite{Il'in_2021_Quantum} and thermal  adiabatic theorem \cite{il'in2020adiabatic} \label{appendix B}}

While proving the bounds presented in this paper, we have employed and improved the technique used previously in refs. \cite{il'in2020adiabatic,Il'in_2021_Quantum}. The improvements allow us to tighten the results of this prior work.

The main improvement is the usage of the Lemma from Section \ref{subsect: Lemma}. A function
\be
f_{E\,E'}^\beta = \frac{e^{-\frac{\beta}{2}E}-e^{-\frac{\beta}{2}E'}}{\beta(E-E')/2}
\ee
plays an important role in refs. \cite{il'in2020adiabatic,Il'in_2021_Quantum}. It has been bounded  in \cite{il'in2020adiabatic,Il'in_2021_Quantum}
by the inequality
\be\label{f_weak}
 \left(f_{E\,E'}^\beta\right)^2\leqslant e^{-\beta E}+e^{-\beta E'}.
\ee
In fact, according to the Lemma \eqref{exp3}, this inequality can be replaced by a tighter one,
\be\label{f_xi}
 \left(f_{E\,E'}^\beta\right)^2\leqslant \frac{e^{-\beta E}+e^{-\beta E'}}{2}.
\ee
In effect, the resulting inequalities in  refs. \cite{il'in2020adiabatic,Il'in_2021_Quantum}   turns out to be twice tighter as the original ones.

A second improvement is the usage of the Cauchy inequality in the integral form \eqref{Cauchy} instead of its weaker versions in refs. \cite{il'in2020adiabatic,Il'in_2021_Quantum}.
\begin{multline}\label{Cauchy}
\left|\int_0^t \tr \left(A_{t'}^\dag \, B_{t'}\right) dt' \right|^2
\leqslant \int_0^t \tr\left(A^\dag_{t'} \, A_{t'}\right)dt'\,~\, \times \\
\times \int_0^t\tr\left(B^\dag_{t'} \, B_{t'}\right)dt'
\end{multline}
As a result, one replaces the adiabatic condition (S43) in \cite{il'in2020adiabatic} by
\begin{multline}
D_t\leqslant \frac{\omega\beta}{2} \,\Bigg(\frac{1}{\mu_{\omega t}}\|V_{\omega t}\|+\sqrt{\omega t\int_{0}^{\omega t}\frac{1}{\mu_{s'}^2}\|\partial_{t'}V_{s'}\|^2ds'}+\\
+\sqrt{\omega t\int_{0}^{\omega t}\frac{\nu_{s'}^2}{\mu_{s'}^2}\|V_{s'}\|^2ds'}+\\
+ \sqrt2 \, \sqrt{\int_{0}^{\omega t}\frac{1}{\mu_{s'}^2}\|V_{s'}\|^2ds'} \, \sqrt{\int_{0}^{\omega t}\|V_{s'}\|^2ds'}\Bigg).
\end{multline}
The notations here are defined in ref. \cite{il'in2020adiabatic}.  The adiabatic condition (13) in \cite{il'in2020adiabatic} is improved accordingly.

Analogously, one replaces the QSL~(30) from ref. \cite{Il'in_2021_Quantum} by its stronger version:
\begin{align}\label{QSL thermal}
D_t \leqslant & \frac{\beta}2 \, \sqrt{
t\int_0^t dt' \, \langle -[H_0,V_{t'}]^2 \rangle_\beta
}.
\end{align}
The reader is referred to ref. \cite{Il'in_2021_Quantum} for the explanation of the setup and notations in this formula. The QSLs (8) and (21) in ref.  \cite{Il'in_2021_Quantum} are modified accordingly.

\bibliography{references,C:/D/Work/QM/JabRef/bibliography,C:/D/Work/QM/Bibs/LZ_and_adiabaticity,C:/D/Work/QM/Bibs/AQC,C:/D/Work/QM/Bibs/QIP,C:/D/Work/QM/Bibs/QSL,C:/D/Work/QM/Bibs/1D,C:/D/Work/QM/Bibs/dynamically_integrable}
\bibliographystyle{apsrev4-1}

\end{document}